\RequirePackage{lineno}
\documentclass[aps,prd,twocolumn,superscriptaddress,groupedaddress,floatfix,letter]{revtex4}

\usepackage{graphicx}% Include figure files
\usepackage{dcolumn}% Align table columns on decimal point
\usepackage{bm}% bold math
\usepackage{amsmath,amssymb}   % for math
\usepackage{subfigure}
\usepackage{multirow}
\usepackage{afterpage}
\usepackage{url}
\usepackage{latexsym}
\usepackage{lscape}
\usepackage{subfigure}
\usepackage{setspace}

\begin{document}

%\hspace{4.8in} \mbox{FERMILAB-PUB-19-253-E }

\noindent%
\mbox{FERMILAB-PUB-19-253-E\hspace{15mm}{\em Published in Phys. Rev. D as 
DOI: 10.1103/PhysRevD.100.012005}}

\title{Properties of    {\boldmath  $Z_c^{\pm}(3900)$} produced in {\boldmath  $p \bar p$} collisions }

\affiliation{LAFEX, Centro Brasileiro de Pesquisas F\'{i}sicas, Rio de Janeiro, RJ 22290, Brazil}
\affiliation{Universidade do Estado do Rio de Janeiro, Rio de Janeiro, RJ 20550, Brazil}
\affiliation{Universidade Federal do ABC, Santo Andr\'e, SP 09210, Brazil}
\affiliation{University of Science and Technology of China, Hefei 230026, People's Republic of China}
\affiliation{Universidad de los Andes, Bogot\'a, 111711, Colombia}
\affiliation{Charles University, Faculty of Mathematics and Physics, Center for Particle Physics, 116 36 Prague 1, Czech Republic}
\affiliation{Czech Technical University in Prague, 116 36 Prague 6, Czech Republic}
\affiliation{Institute of Physics, Academy of Sciences of the Czech Republic, 182 21 Prague, Czech Republic}
\affiliation{Universidad San Francisco de Quito, Quito 170157, Ecuador}
\affiliation{LPC, Universit\'e Blaise Pascal, CNRS/IN2P3, Clermont, F-63178 Aubi\`ere Cedex, France}
\affiliation{LPSC, Universit\'e Joseph Fourier Grenoble 1, CNRS/IN2P3, Institut National Polytechnique de Grenoble, F-38026 Grenoble Cedex, France}
\affiliation{CPPM, Aix-Marseille Universit\'e, CNRS/IN2P3, F-13288 Marseille Cedex 09, France}
\affiliation{LAL, Univ. Paris-Sud, CNRS/IN2P3, Universit\'e Paris-Saclay, F-91898 Orsay Cedex, France}
\affiliation{LPNHE, Universit\'es Paris VI and VII, CNRS/IN2P3, F-75005 Paris, France}
\affiliation{CEA Saclay, Irfu, SPP, F-91191 Gif-Sur-Yvette Cedex, France}
\affiliation{IPHC, Universit\'e de Strasbourg, CNRS/IN2P3, F-67037 Strasbourg, France}
\affiliation{IPNL, Universit\'e Lyon 1, CNRS/IN2P3, F-69622 Villeurbanne Cedex, France and Universit\'e de Lyon, F-69361 Lyon CEDEX 07, France}
\affiliation{III. Physikalisches Institut A, RWTH Aachen University, 52056 Aachen, Germany}
\affiliation{Physikalisches Institut, Universit\"at Freiburg, 79085 Freiburg, Germany}
\affiliation{II. Physikalisches Institut, Georg-August-Universit\"at G\"ottingen, 37073 G\"ottingen, Germany}
\affiliation{Institut f\"ur Physik, Universit\"at Mainz, 55099 Mainz, Germany}
\affiliation{Ludwig-Maximilians-Universit\"at M\"unchen, 80539 M\"unchen, Germany}
\affiliation{Panjab University, Chandigarh 160014, India}
\affiliation{Delhi University, Delhi-110 007, India}
\affiliation{Tata Institute of Fundamental Research, Mumbai-400 005, India}
\affiliation{University College Dublin, Dublin 4, Ireland}
\affiliation{Korea Detector Laboratory, Korea University, Seoul, 02841, Korea}
\affiliation{CINVESTAV, Mexico City 07360, Mexico}
\affiliation{Nikhef, Science Park, 1098 XG Amsterdam, the Netherlands}
\affiliation{Radboud University Nijmegen, 6525 AJ Nijmegen, the Netherlands}
\affiliation{Joint Institute for Nuclear Research, Dubna 141980, Russia}
\affiliation{Institute for Theoretical and Experimental Physics, Moscow 117259, Russia}
\affiliation{Moscow State University, Moscow 119991, Russia}
\affiliation{Institute for High Energy Physics, Protvino, Moscow region 142281, Russia}
\affiliation{Petersburg Nuclear Physics Institute, St. Petersburg 188300, Russia}
\affiliation{Instituci\'{o} Catalana de Recerca i Estudis Avan\c{c}ats (ICREA) and Institut de F\'{i}sica d'Altes Energies (IFAE), 08193 Bellaterra (Barcelona), Spain}
\affiliation{Uppsala University, 751 05 Uppsala, Sweden}
\affiliation{Taras Shevchenko National University of Kyiv, Kiev, 01601, Ukraine}
\affiliation{Lancaster University, Lancaster LA1 4YB, United Kingdom}
\affiliation{Imperial College London, London SW7 2AZ, United Kingdom}
\affiliation{The University of Manchester, Manchester M13 9PL, United Kingdom}
\affiliation{University of Arizona, Tucson, Arizona 85721, USA}
\affiliation{University of California Riverside, Riverside, California 92521, USA}
\affiliation{Florida State University, Tallahassee, Florida 32306, USA}
\affiliation{Fermi National Accelerator Laboratory, Batavia, Illinois 60510, USA}
\affiliation{University of Illinois at Chicago, Chicago, Illinois 60607, USA}
\affiliation{Northern Illinois University, DeKalb, Illinois 60115, USA}
\affiliation{Northwestern University, Evanston, Illinois 60208, USA}
\affiliation{Indiana University, Bloomington, Indiana 47405, USA}
\affiliation{Purdue University Calumet, Hammond, Indiana 46323, USA}
\affiliation{University of Notre Dame, Notre Dame, Indiana 46556, USA}
\affiliation{Iowa State University, Ames, Iowa 50011, USA}
\affiliation{University of Kansas, Lawrence, Kansas 66045, USA}
\affiliation{Louisiana Tech University, Ruston, Louisiana 71272, USA}
\affiliation{Northeastern University, Boston, Massachusetts 02115, USA}
\affiliation{University of Michigan, Ann Arbor, Michigan 48109, USA}
\affiliation{Michigan State University, East Lansing, Michigan 48824, USA}
\affiliation{University of Mississippi, University, Mississippi 38677, USA}
\affiliation{University of Nebraska, Lincoln, Nebraska 68588, USA}
\affiliation{Rutgers University, Piscataway, New Jersey 08855, USA}
\affiliation{Princeton University, Princeton, New Jersey 08544, USA}
\affiliation{State University of New York, Buffalo, New York 14260, USA}
\affiliation{University of Rochester, Rochester, New York 14627, USA}
\affiliation{State University of New York, Stony Brook, New York 11794, USA}
\affiliation{Brookhaven National Laboratory, Upton, New York 11973, USA}
\affiliation{Langston University, Langston, Oklahoma 73050, USA}
\affiliation{University of Oklahoma, Norman, Oklahoma 73019, USA}
\affiliation{Oklahoma State University, Stillwater, Oklahoma 74078, USA}
\affiliation{Oregon State University, Corvallis, Oregon 97331, USA}
\affiliation{Brown University, Providence, Rhode Island 02912, USA}
\affiliation{University of Texas, Arlington, Texas 76019, USA}
\affiliation{Southern Methodist University, Dallas, Texas 75275, USA}
\affiliation{Rice University, Houston, Texas 77005, USA}
\affiliation{University of Virginia, Charlottesville, Virginia 22904, USA}
\affiliation{University of Washington, Seattle, Washington 98195, USA}
\author{V.M.~Abazov} \affiliation{Joint Institute for Nuclear Research, Dubna 141980, Russia}
\author{B.~Abbott} \affiliation{University of Oklahoma, Norman, Oklahoma 73019, USA}
\author{B.S.~Acharya} \affiliation{Tata Institute of Fundamental Research, Mumbai-400 005, India}
\author{M.~Adams} \affiliation{University of Illinois at Chicago, Chicago, Illinois 60607, USA}
\author{T.~Adams} \affiliation{Florida State University, Tallahassee, Florida 32306, USA}
\author{J.P.~Agnew} \affiliation{The University of Manchester, Manchester M13 9PL, United Kingdom}
\author{G.D.~Alexeev} \affiliation{Joint Institute for Nuclear Research, Dubna 141980, Russia}
\author{G.~Alkhazov} \affiliation{Petersburg Nuclear Physics Institute, St. Petersburg 188300, Russia}
\author{A.~Alton$^{a}$} \affiliation{University of Michigan, Ann Arbor, Michigan 48109, USA}
\author{A.~Askew} \affiliation{Florida State University, Tallahassee, Florida 32306, USA}
% ...
%
% visitor_addresses.tex                       18 June 2018
%  available symbols are:
%  $\ast, \dag, \ddag, \S, \P, $\|$, $\ast\ast$, \dag\dag, \ddag\ddag ,\#
%
\collaboration{The D0 Collaboration\footnote{with visitors from
%{alton}
$^{a}$Augustana College, Sioux Falls, SD 57197, USA,
%{burdin}
$^{b}$The University of Liverpool, Liverpool L69 3BX, UK,
%{grohsjean,deterre}
$^{c}$Deutshes Elektronen-Synchrotron (DESY), Notkestrasse 85, Germany,
%{de la cruz-burelo}
$^{d}$CONACyT, M-03940 Mexico City, Mexico,
%{partridge}
$^{e}$SLAC, Menlo Park, CA 94025, USA,
%{hesketh}
$^{f}$University College London, London WC1E 6BT, UK,
%{luna-garcia}
$^{g}$Centro de Investigacion en Computacion - IPN, CP 07738 Mexico City, Mexico,
%{santos}
$^{h}$Universidade Estadual Paulista, S\~ao Paulo, SP 01140, Brazil,
%{meyer}
$^{i}$Karlsruher Institut f\"ur Technologie (KIT) - Steinbuch Centre for Computing (SCC),
D-76128 Karlsruhe, Germany,
%{patwa}
$^{j}$Office of Science, U.S. Department of Energy, Washington, D.C. 20585, USA,
%{cooke}
%$^{k}$American Association for the Advancement of Science, Washington, D.C. 20005, USA,
%{borysova}
$^{l}$Kiev Institute for Nuclear Research (KINR), Kyiv 03680, Ukraine,
%{jabeen}
$^{m}$University of Maryland, College Park, MD 20742, USA,
%{williams}
$^{n}$European Orgnaization for Nuclear Research (CERN), CH-1211 Geneva, Switzerland,
%{Jung}
$^{o}$Purdue University, West Lafayette, IN 47907, USA,
%{Zivkovic}
$^{p}$Institute of Physics, Belgrade, Belgrade, Serbia,
and
%{Drutskoy}
$^{q}$P.N. Lebedev Physical Institute of the Russian Academy of Sciences, 119991, Moscow, Russia.
%{montgomery}
%$^{?}$Thomas Jefferson National Accelerator Facility, Newport News, VA 23606, USA,
%{falkowski}
%$^{?}$Laboratoire de Physique Theorique, F-91405 Orsay CEDEX, FR,
%{hooper,kozminski}
%$^{?}$}Visitor from Lewis University, Romeoville, IL 60446, USA.
%{weber}
%$^{?}$Universit{\"a}t Bern, CH-3012 Bern, Switzerland.
%{deceased}
%{peng, lipaev, cihangir}
$^{\ddag}$Deceased.
}} \noaffiliation
\vskip 0.25cm

\date{May 31, 2019}

\begin{abstract}
We study the production of  the exotic charged charmonium-like state
$Z_c^{\pm}(3900)$ in $p \bar p$ collisions through  the sequential  process
$\psi(4260) \rightarrow Z_c^{\pm}(3900) \pi^{\mp}$,
 $Z_c^{\pm}(3900) \rightarrow J/\psi \pi^{\pm}$.
Using the subsample of candidates  originating from
 semi-inclusive weak decays of $b$-flavored hadrons, we measure the
invariant mass and natural width 
to be $M=3902.6 ^{+5.2}_{-5.0}{\rm \thinspace (stat)}^{+3.3}_{-1.4}{\rm \thinspace (syst)}$~MeV and
 $\Gamma=32 ^{+28}_{-21}{\rm \thinspace (stat)} ^{+26}_{-7}{\rm \thinspace (syst)} $~MeV, 
respectively.
We search for prompt production 
of the  $Z_c^{\pm}(3900)$ through  the same sequential  process.
No significant signal is observed, and we set an upper limit of 0.70
at the  95\% credibility level 
 on the ratio
of prompt production to the production  via $b$-hadron decays. 
The study is based on  $10.4~\rm{fb^{-1}}$ of $p \overline p $ collision
data  collected by the  D0 experiment at  the Fermilab Tevatron collider.
\end{abstract}

\maketitle

\section{\label{sec:intro}Introduction}

In high-energy  hadron collisions, charmonium  is known to be produced
both promptly in QCD processes and  non-promptly in $b$-hadron decays,
with well  measured rates. For both $J/\psi$ and $\psi(2S)$ mesons
the non-prompt fraction  increases with transverse momentum
but  prompt production dominates  in most of the  studied $p_T$ range~\cite{fb}. 
 
Much less information exists about the hadronic production of exotic multiquark
states containing a charm quark and antiquark.
The $X(3872)$ -- the most extensively studied exotic meson --
is produced copiously in prompt
  $p \overline p$ interactions at $\sqrt{s}=1.96$~TeV~\cite{d03872},
and in $pp$ collisions at  $\sqrt{s}=7$~TeV~\cite{7tev3872} and 
$\sqrt{s}=8$~TeV~\cite{8tev3872}.
The fraction  of the inclusive production rate of the  $X(3872)$ mesons
originating  from decays of $b$-flavored hadrons ($H_b$)  
 is found to be approximately  0.3~\cite{7tev3872,8tev3872}, independent of $p_T$.
Evidence for prompt production of the $X(4140)$, another exotic candidate,
 was also reported by D0~\cite{d04140}. 
The large prompt production rate of the $X(3872)$ has often been used as
an argument against its identification as a weakly bound charm-meson molecule; see
Ref.~\cite{bra3872} for the latest discussion.

 In Ref.~\cite{zc1}, the D0 Collaboration presented the first evidence for
production of the manifestly exotic charmonium-like state $Z_c^{\pm}(3900)$ in semi-inclusive weak
decays of $b$-flavored hadrons in events containing a non-prompt $J/\psi$ and a pair of
 oppositely charged particles,  assumed to be pions. That analysis considered the mass range
 $4.1<M(J/\psi \pi^+ \pi^-)<4.7$~GeV that includes the $\psi(4260)$ state:
$H_b \rightarrow  \psi(4260) + {\rm anything}$,  $\psi(4260) \rightarrow Z_c^{\pm}(3900) \pi^{\mp}$, 
 $Z_c^{\pm}(3900) \rightarrow J/\psi \pi^{\pm}$. 
This article presents  an extension of  that study to a search for prompt production 
of the  $Z_c^{\pm}(3900)$ through  the sequential  process
$\psi(4260) \rightarrow Z_c^{\pm}(3900) \pi^{\mp}$,
 $Z_c^{\pm}(3900) \rightarrow J/\psi \pi^{\pm}$.
The event sample used in this analysis is approximately 50\% larger than in  Ref.~\cite{zc1}
due to the use of an extended track finding algorithm optimized for reconstructing low-$p_T$ tracks. 

\section{\label{sec:event}The D0 detector, event reconstruction, and selection}

%\subsection{D0 detector}
The D0 detector has a central tracking system consisting of a silicon
microstrip tracker  and a central fiber tracker, both located within a
1.9~T superconducting solenoidal magnet~\cite{d0det, layer0}. A muon
system, covering $|\eta|<2$~\cite{eta}, consists of a layer of tracking
detectors and scintillation trigger counters in front of a central and two forward 1.8~T iron toroidal
magnets, followed by two similar layers after the toroids~\cite{run2muon}. 
Events used in this analysis are collected with both single-muon and dimuon triggers.
Single-muon triggers require a coincidence of signals in trigger elements inside and outside
the toroidal magnets.
 All dimuon triggers  require at least one muon to have track segments  after the toroid;  
muons in the forward region are always required to penetrate the   toroid.  

The minimum muon transverse momentum is 1.5 GeV.
No minimum $p_T$ requirement is applied to the muon pair,
but the effective threshold is approximately 4 GeV due to the requirement for muons
to penetrate the toroids,  and the average  value for 
accepted events is 10 GeV.

In  $p \overline p$ collisions the $J/\psi$ is produced
promptly, either directly or in strong decays of higher-mass charmonium states,
or non-promptly  in $b$-hadron decays.
Prompt mesons have a decay vertex consistent with the interaction point
while those from the $b$ decays are displaced on average by   $\mathcal O$(1~mm)
as a result of the long $b$-hadron lifetime.

\begin{figure}[htb]
\includegraphics[scale=0.42]{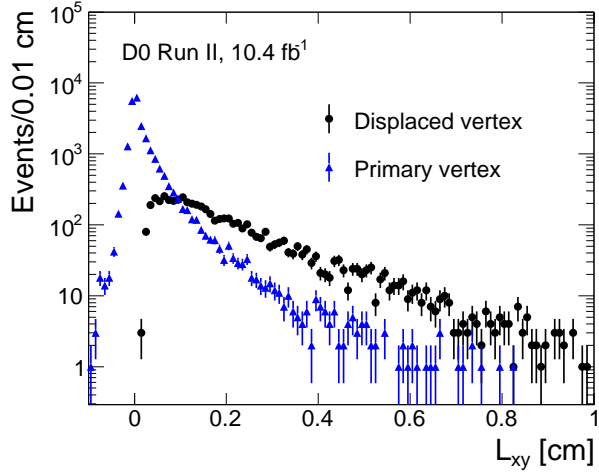}
\caption{\label{fig:Lxy} 
The $J/\psi \pi^+ \pi^-$ decay length  in the transverse plane for events in the range
 $4.2<M(J/\psi \pi^+ \pi^-)<4.3$~GeV. The black filled circles show the distribution of
events that satisfy the  criteria for a displaced vertex. This subsample
constitutes about 2/3 of the nonprompt events. The distribution marked with blue triangles
includes the prompt production and the remaining 1/3 of the nonprompt events.
}
\end{figure}

We reconstruct $J/\psi \rightarrow \mu^+ \mu^-$ decay candidates accompanied
by a pair of charged particles, assumed to be pions, 
 with opposite charges and  with $p_T>0.7$~GeV.
 We perform  a   kinematic fit  under the hypothesis that the muons come from
the $J/\psi$ and that the $J/\psi$ and the two particles  originate from 
the same   space point.
In the fit,  the dimuon invariant mass is constrained to the world-average value 
of the  $J/\psi$ meson  mass~\cite{pdg}.
The  track parameters ($p_T$, position and direction in 3D) are readjusted
according to the  fit and
are used in the calculation of the system's transverse
decay-path vector ${\vec L_{\rm xy}}$, 
the  invariant mass $M(J/\psi \pi^+ \pi^-)$, and the  masses of the two $J/\psi \pi$ subsystems. 
Following Refs.~\cite{belle2013} and \cite{bes2013}, 
 we  select the larger mass combination as a $Z_c^{\pm}(3900)$ candidate's mass.

We  select events in the $M(J/\psi \pi^+ \pi^-)$  range
 4.1--4.7~GeV that includes the $\psi(4260)$  and excludes 
fully reconstructed
decays of $b$ hadrons to final states  $J/\psi h_1^+ h_2^-$
where $h_1$ and $h_2$ stand for a pion, a kaon, or  a proton.
We divide the data into two non-overlapping samples: events
with a displaced vertex, selected as in 
Ref.~\cite{zc1}, and a complementary sample of ``primary vertex'' events.
The criteria for the displaced vertex category are:
the  vertex of the $J/\psi$ and  the highest $p_T$ track  is  required to be  displaced in the transverse plane
from the $p\bar{p}$  interaction vertex by at least 5$\sigma$, the significance of the impact parameter 
in the transverse plane  ({\sl IP})~\cite{ip}
of the leading track  is required to be greater than 2$\sigma$,
the second track's   {\sl IP}  significance is 
required to be greater  than 1$\sigma$, and  the second track's contribution to the $J/\psi$+2 tracks
vertex $\chi^2$ must be  less than 6. The cosine of the  angle in the transverse plane between
 the momentum vector and decay path
of  the  $J/\psi +2$~tracks system is required to be greater
than 0.9.

The sample includes events where the hadronic pair comes from decays
$K^*$$\rightarrow$$K \pi$ or $\phi$$\rightarrow$$KK$.
We remove such events by assuming that one or both of the charged hadrons are kaons and vetoing the mass combinations
$0.81<M(\pi K)<0.97$ GeV  and  $1.01<M(K K)<1.03$ GeV.
We also veto photon conversions by removing events with  $M(\pi^+ \pi^-)<0.35$ GeV.
The decay-length distributions  in the transverse plane for  events in the ``displaced vertex''  and the ``primary vertex''
categories in the mass range $4.2<M(J/\psi \pi^+ \pi^-)<4.3$~GeV  are shown in Fig.~\ref{fig:Lxy}.

\section{\label{sec:fitting}$J/\psi \pi^{\pm}$ mass fits }

We study   the $J/\psi\pi^{\pm}$  system in the vicinity of the  $Z_c^{\pm}(3900)$.
We   perform  a binned maximum-likelihood fit  of  the  $M(J/\psi\pi)$ distribution
 to a sum of a resonant signal  and an incoherent background
in six intervals of $M(J/\psi \pi^+ \pi^-)$:
4.1--4.2~GeV, 4.2--4.3~GeV, 4.3--4.4~GeV, 4.4--4.5~GeV, 4.5--4.6~GeV, and 4.6--4.7~GeV.
The signal is  represented by
the $S$-wave relativistic Breit-Wigner function  convolved with a Gaussian mass resolution. 
The $Z_c^{\pm}(3900)$ mass and width are fixed to the  values for  the 
$J/\psi \pi^{\pm,0}$ channels only (see Ref.~\cite{zcpdg1}):   $M=3893.3\pm2.7$~MeV,
$\Gamma=36.8\pm 6.5$~MeV.
The D0 mass resolution at this mass is $\sigma=17\pm2$~MeV.
In these fits we allow negative values for the signal yield.

For the ``displaced  vertex'' selection, background is mainly due to
weak decays of  $b$ hadrons to a $J/\psi$ paired randomly
with hadrons coming from the same multi-body decay.
For the ``primary vertex'' events, the main background is due to a promptly produced $J/\psi$ combined
with particles produced in the hadronization process.  
In both cases  we use  Chebyshev polynomials of the first kind to represent background.
The fitting range limits  are chosen so as to 
obtain an acceptable fit in a  maximum
range while avoiding areas where the total probability density function goes to zero.
We choose the  order of the Chebyshev polynomial 
to minimize 
the Akaike information test  ($AIC$)~\cite{aic}.
For a fit with $p$ free parameters to a distribution in $n$ bins the $AIC$ is defined as
$AIC = \chi^2 +2p + 2p(p+1)/(n-p-1)$. 
For the displaced-vertex subsample we choose a 4th-order polynomial, and
for the ``primary vertex'' sample  the choice is a 5th-order polynomial.

\section{\label{sec:results}Fit results}

\begin{figure*}[]
\includegraphics[height=6cm,width=0.89\columnwidth]{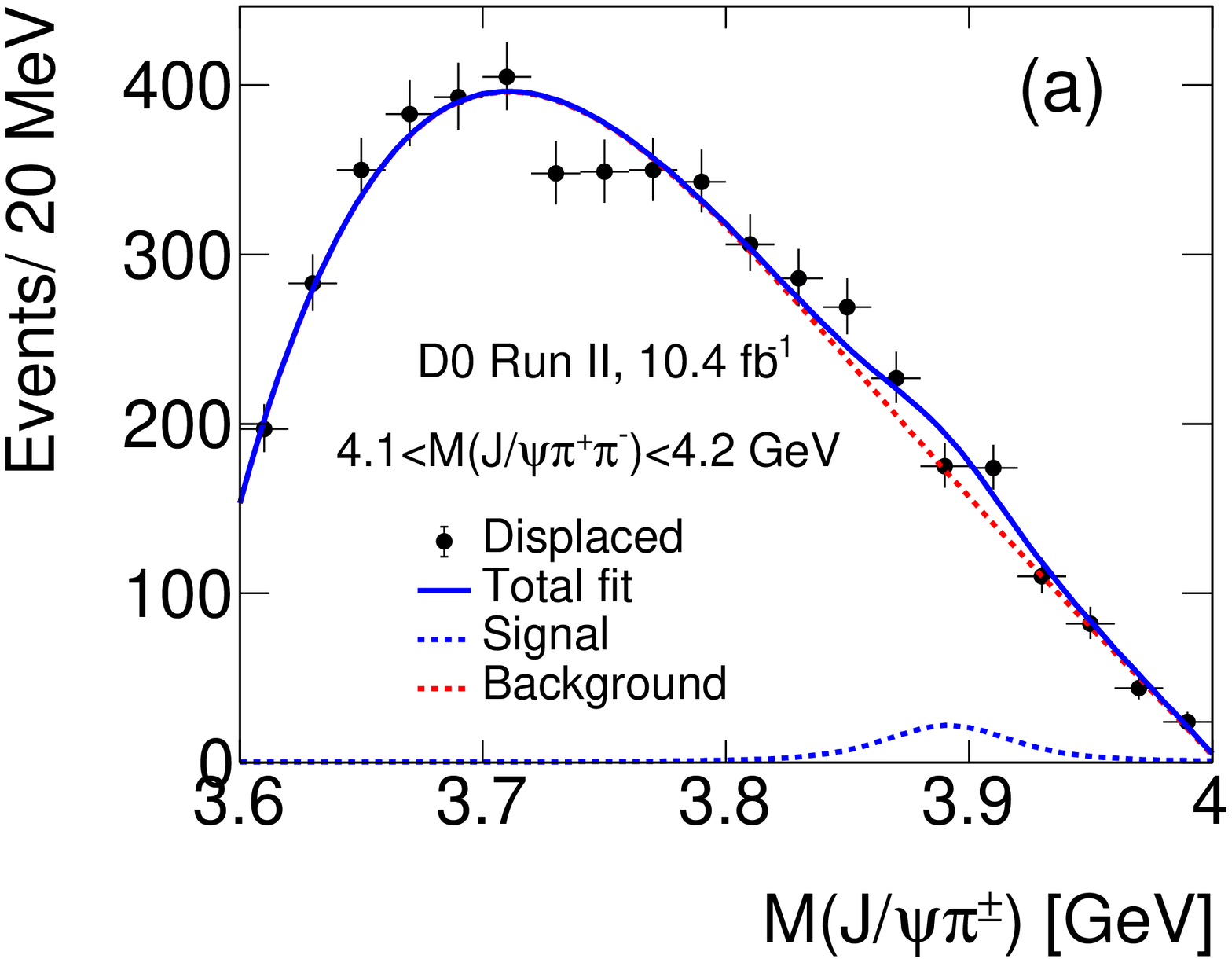}
\includegraphics[height=6.0cm,width=0.89\columnwidth]{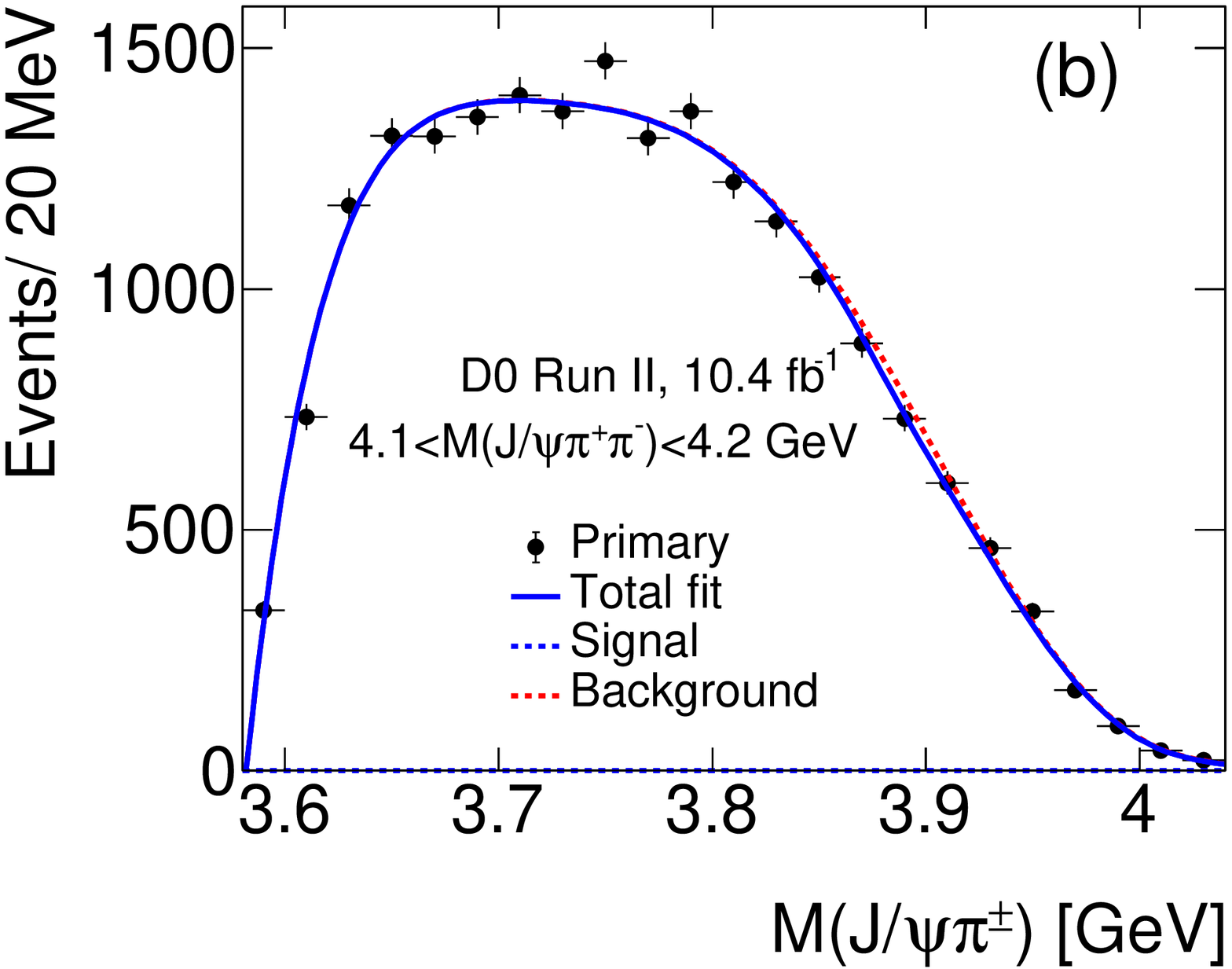}\vspace{5mm}
\includegraphics[height=6.0cm,width=0.89\columnwidth]{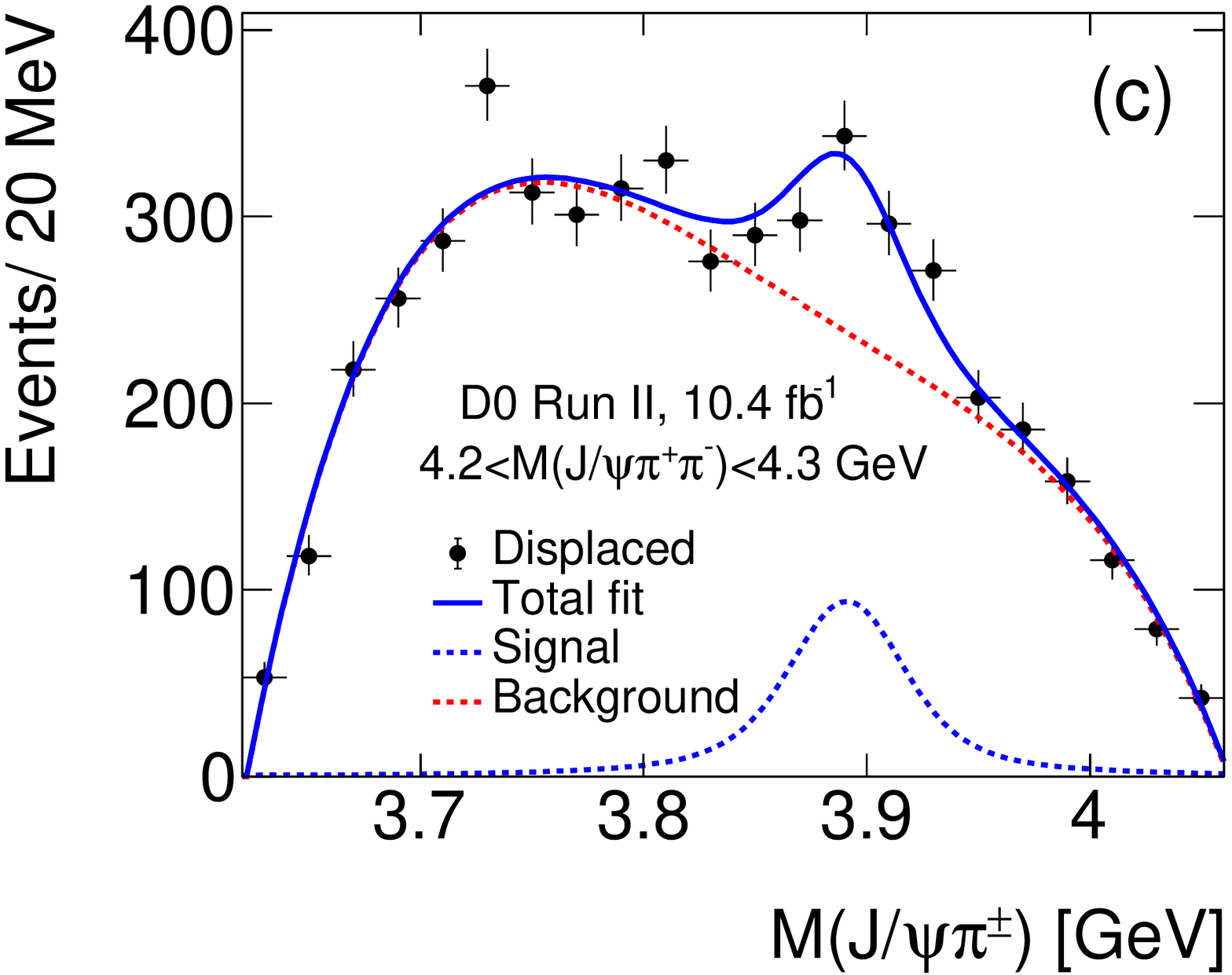}
\includegraphics[height=6.0cm,width=0.89\columnwidth]{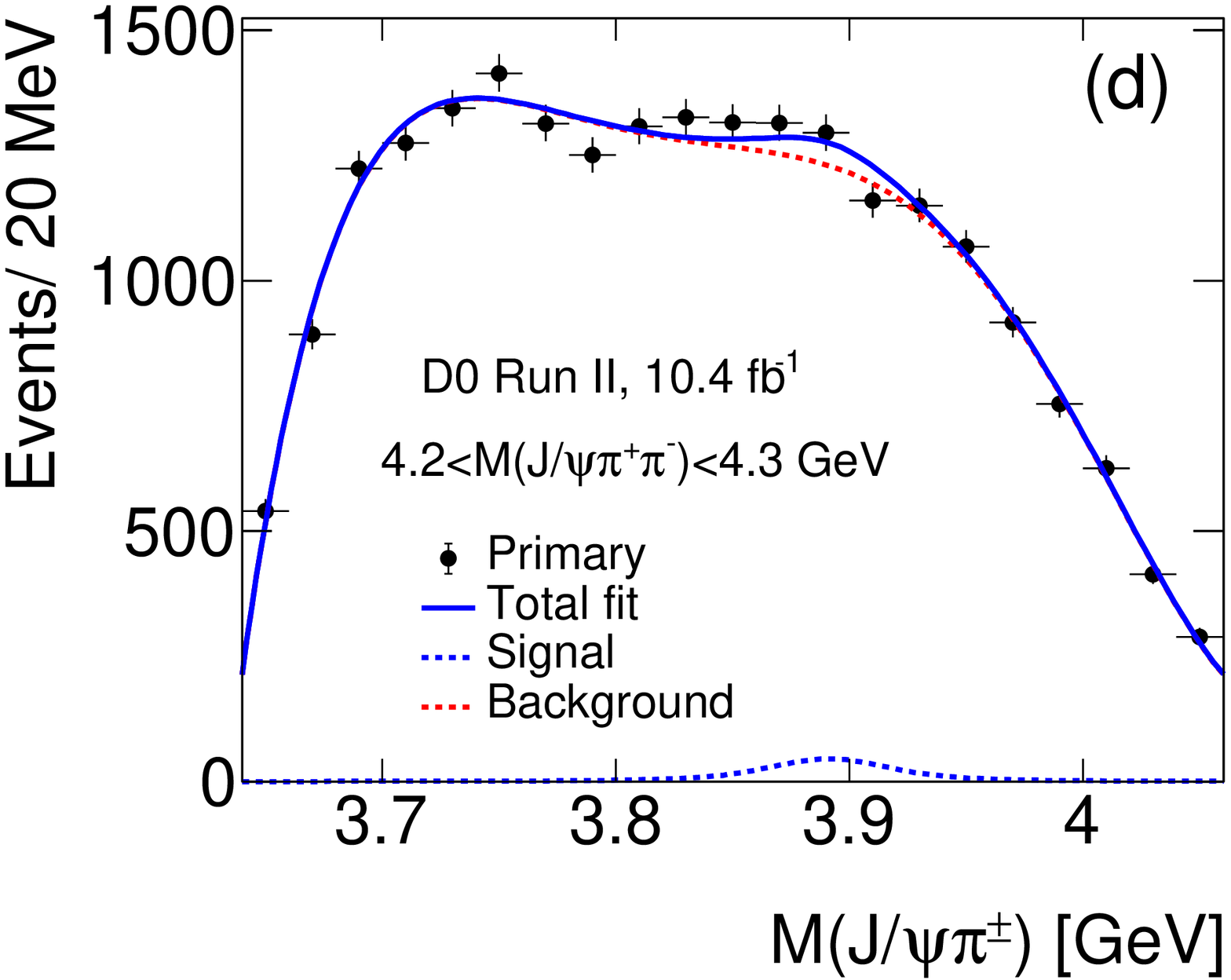}\vspace{5mm}
\includegraphics[height=6.0cm,width=0.89\columnwidth]{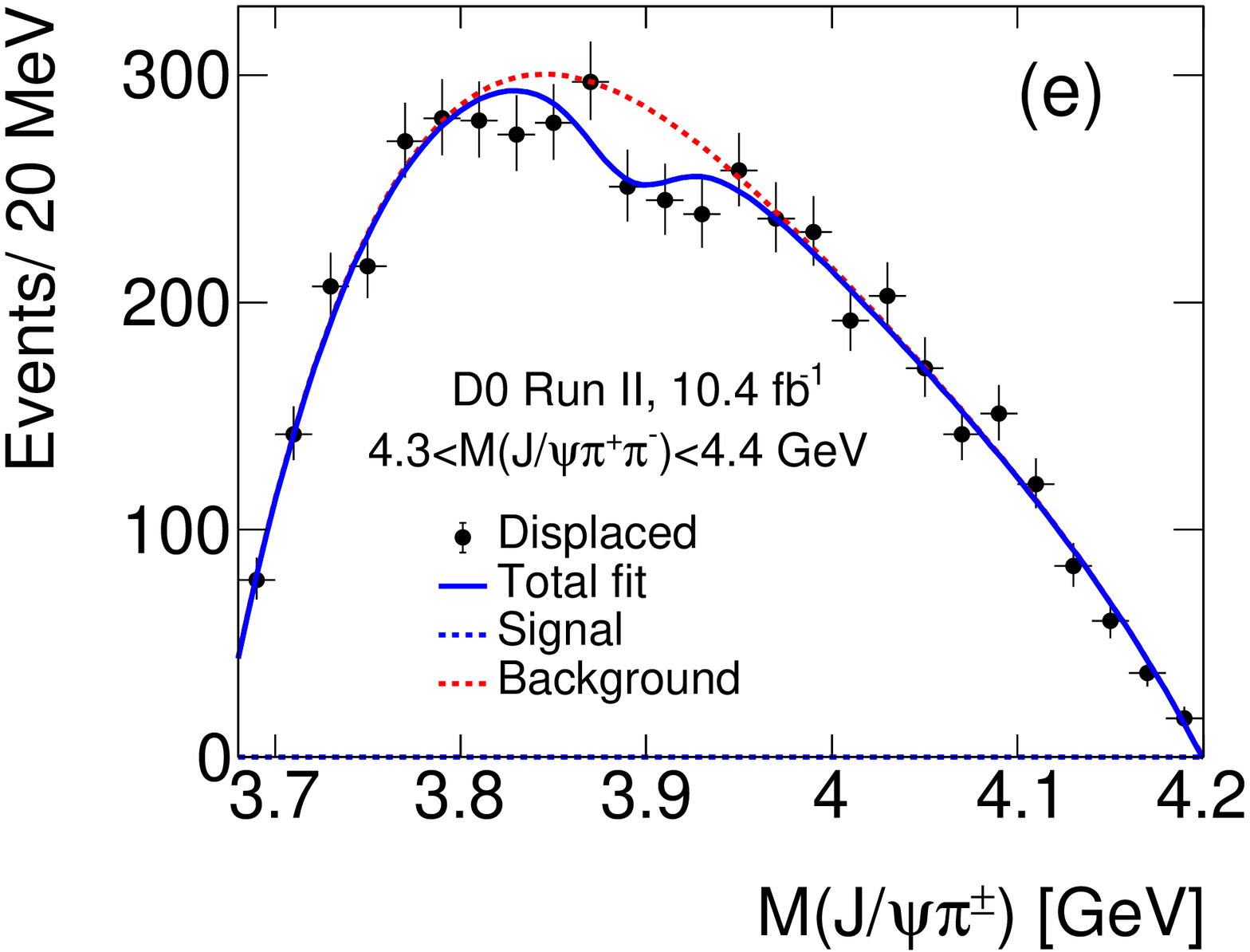}
\includegraphics[height=6.0cm,width=0.89\columnwidth]{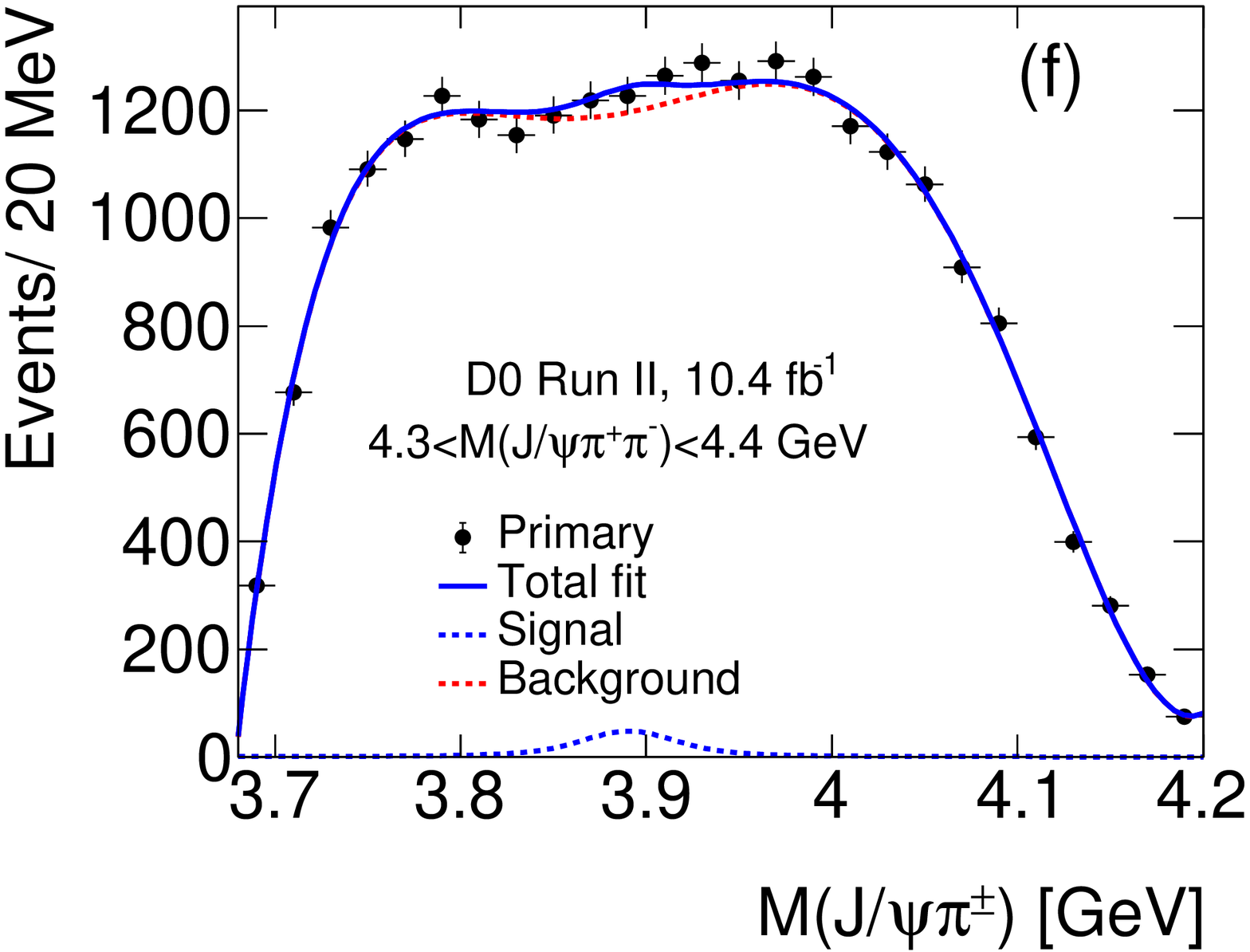}
\caption{\label{fig:npp1} 
The invariant mass distribution of  $J/\psi \pi^{\pm}$ candidates in three intervals of $M(J/\psi \pi^+ \pi^-)$,
from top to bottom 4.1--4.2~GeV, 4.2--4.3~GeV, and 4.3--4.4~GeV.
Left: events with a displaced vertex. Right:  ``primary vertex'' events.
Superimposed are the fits  of a Breit-Wigner signal with fixed mass and width~\cite{zcpdg1} (dashed blue lines),
a Chebyshev polynomial   background (dashed red lines), and their sum (solid blue lines).
}
\end{figure*}

\begin{figure*}[]
\includegraphics[height=6.0cm,width=0.89\columnwidth]{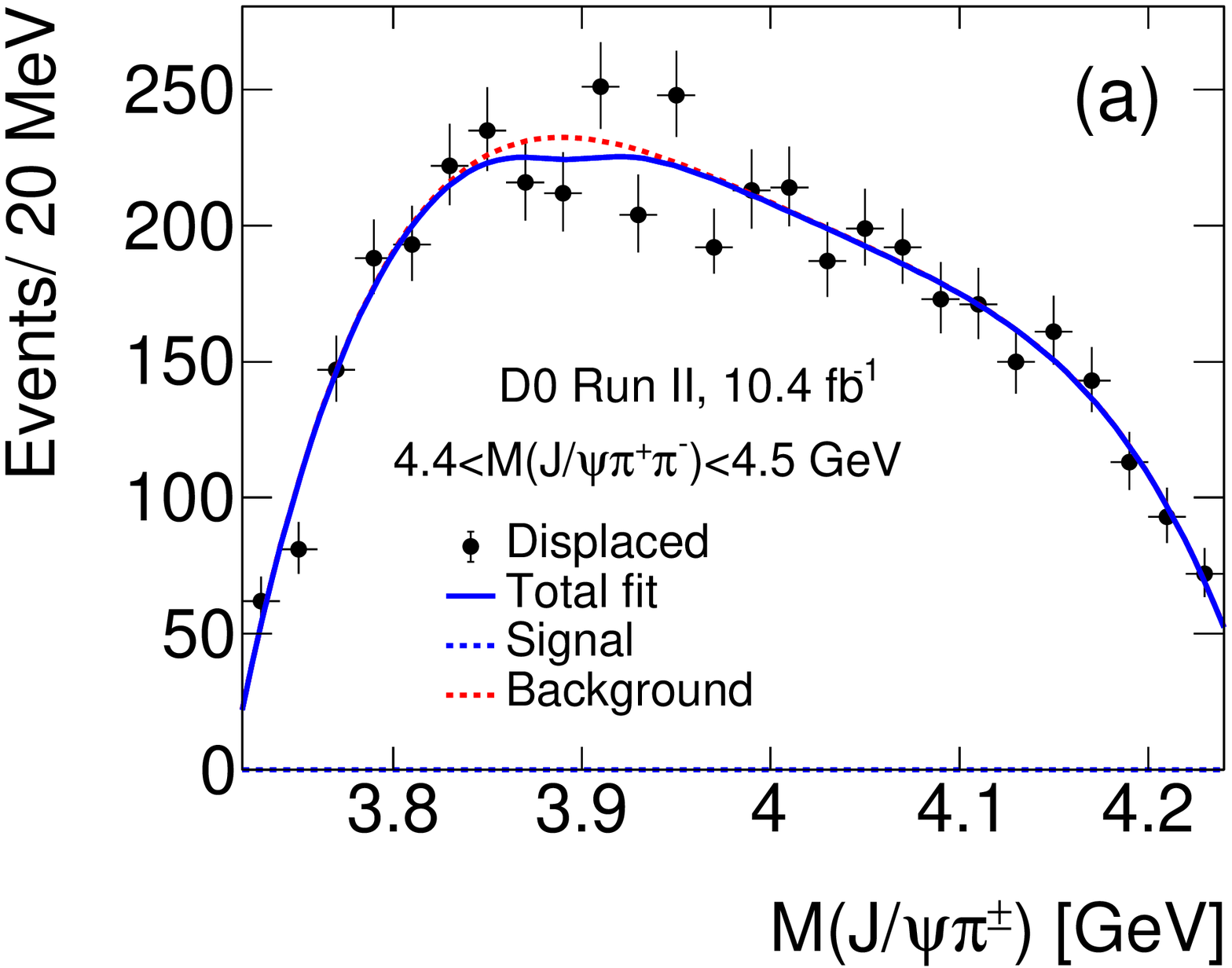}
\includegraphics[height=6.0cm,width=0.89\columnwidth]{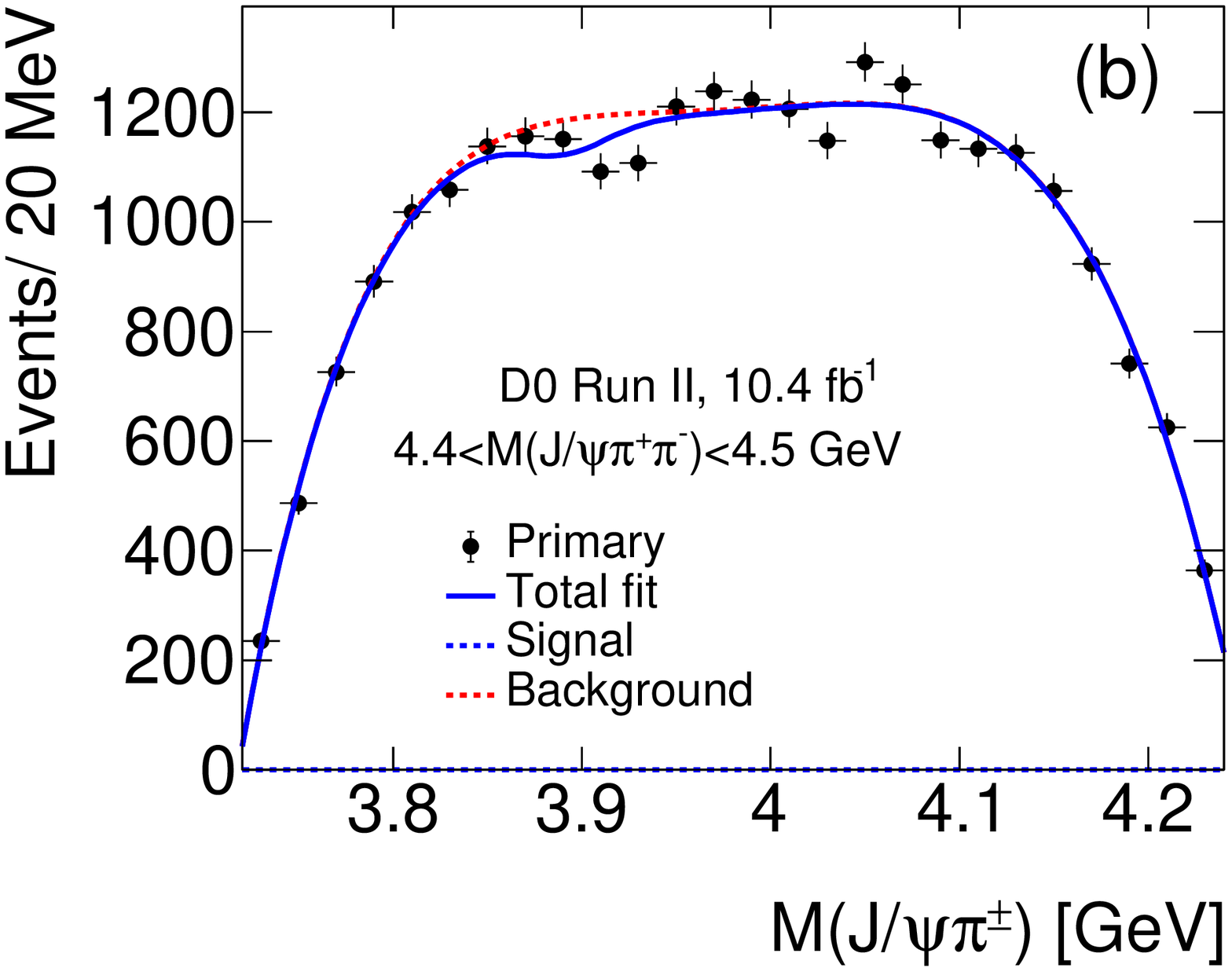}\vspace{5mmm}
\includegraphics[height=6.0cm,width=0.89\columnwidth]{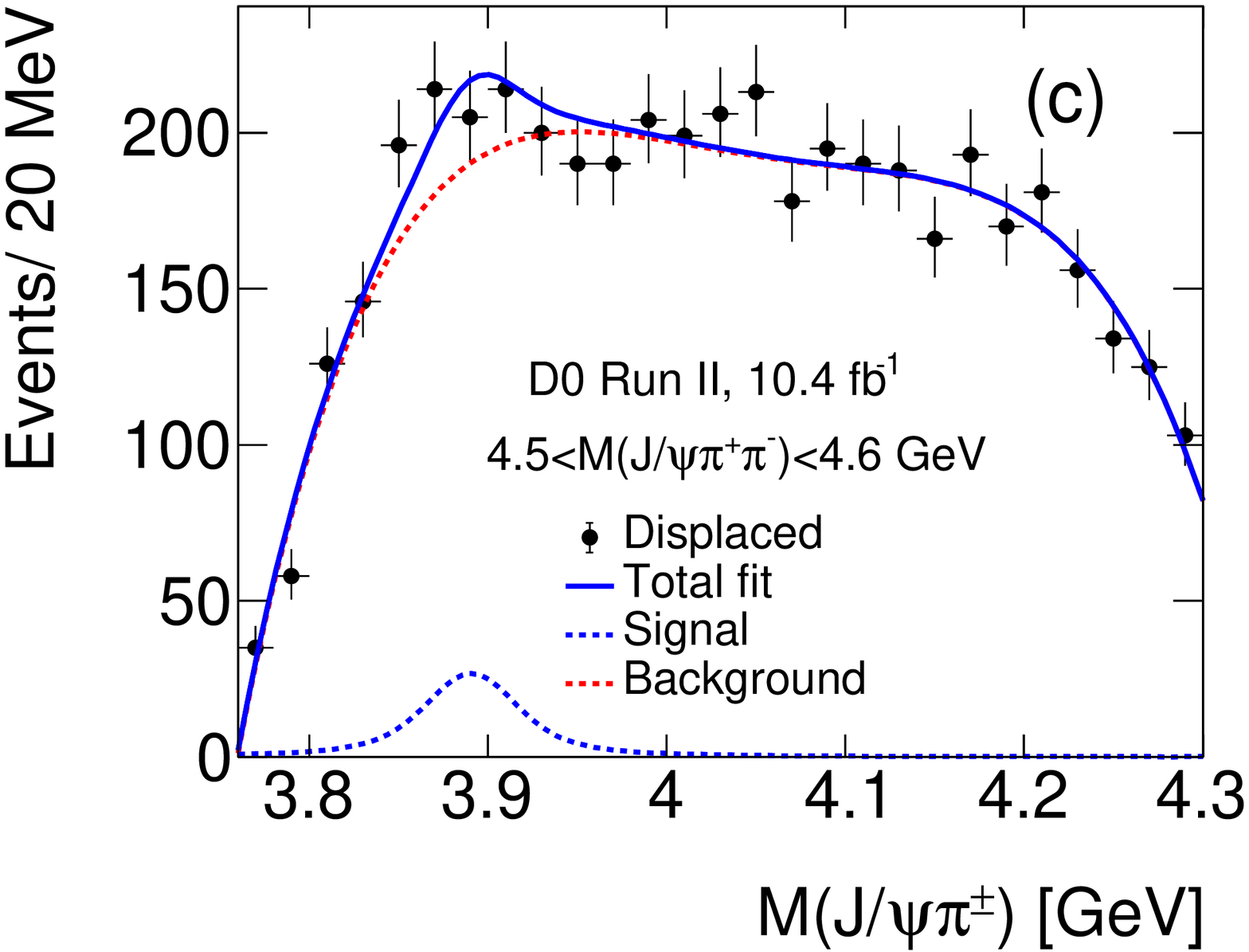}
\includegraphics[height=6.0cm,width=0.89\columnwidth]{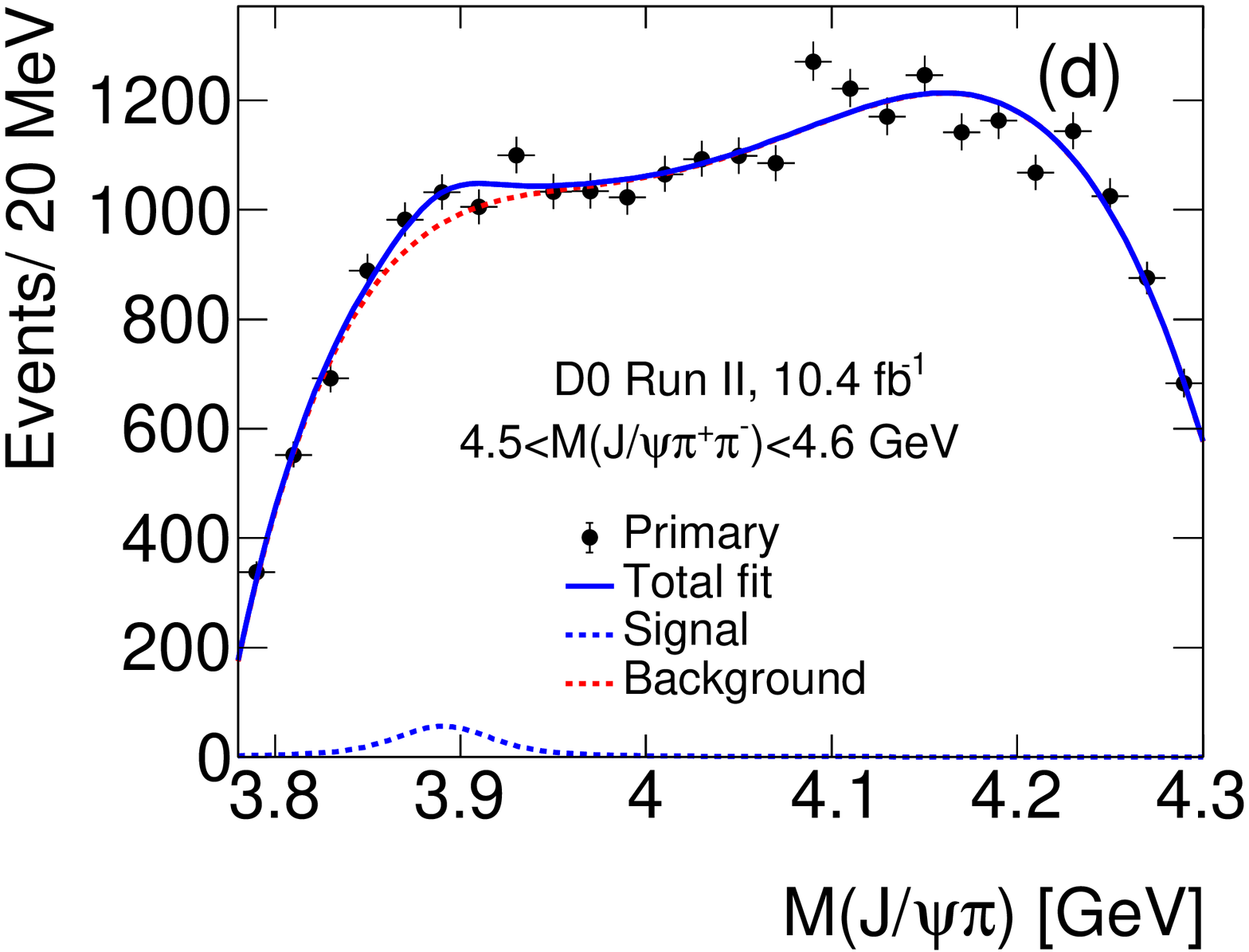}\vspace{5mm}
\includegraphics[height=6.0cm,width=0.89\columnwidth]{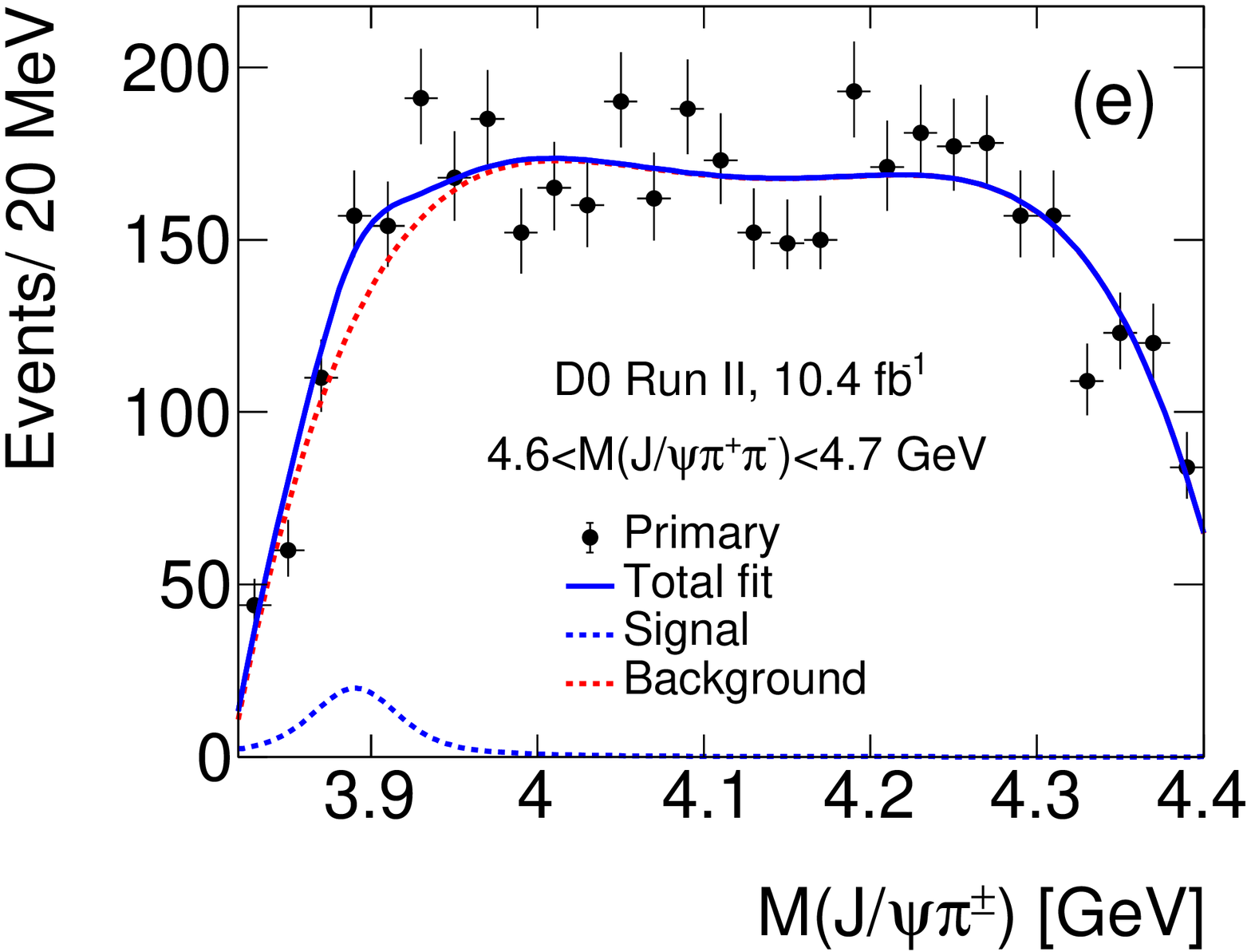}
\includegraphics[height=6.0cm,width=0.89\columnwidth]{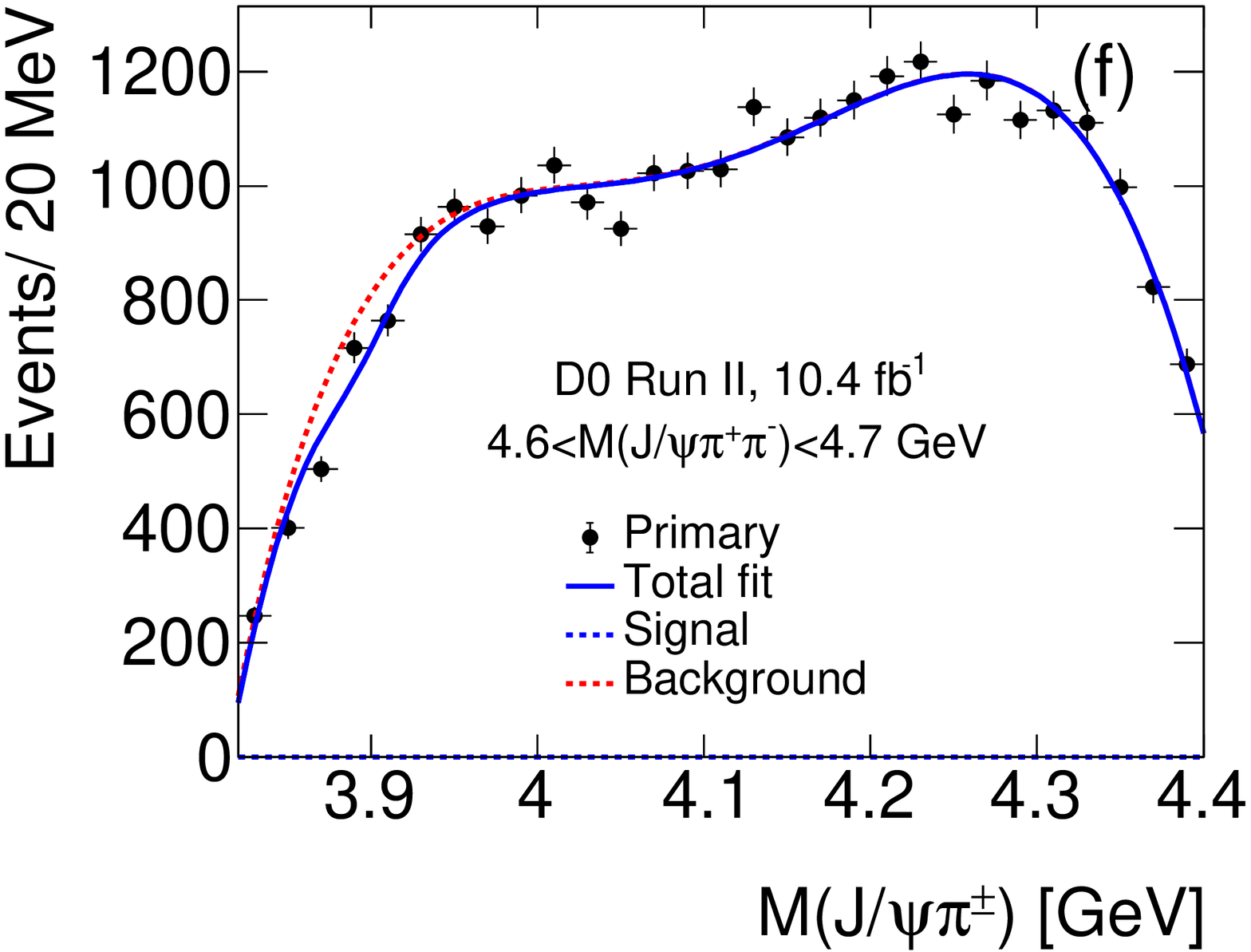}
\caption{\label{fig:npp2} 
The invariant mass distribution of  $J/\psi \pi^{\pm}$ candidates in three intervals of $M(J/\psi \pi^+ \pi^-)$,
from top to bottom 4.4--4.5~GeV,  4.5--4.6~GeV, and 4.6--4.7~GeV.
Left: events with a displaced vertex. Right:  ``primary vertex'' events.
Superimposed are the fits  of a Breit-Wigner signal with fixed mass and width~\cite{zcpdg1}  (dashed blue lines),
a Chebyshev polynomial   background (dashed red lines), and their sum (solid blue lines).
}
\end{figure*}

\begin{figure}[htb]
\includegraphics[width=0.9\columnwidth]{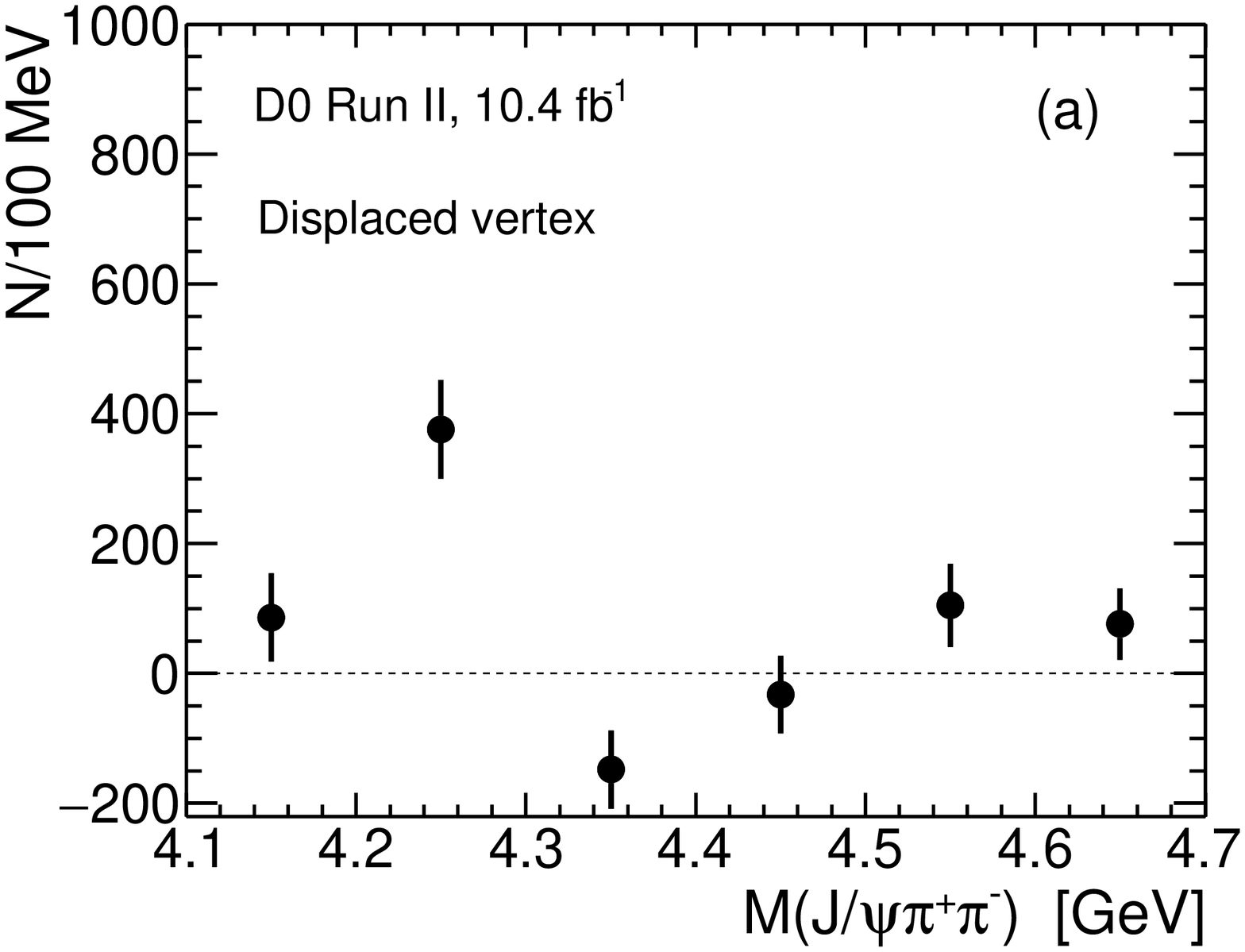}
\includegraphics[width=0.9\columnwidth]{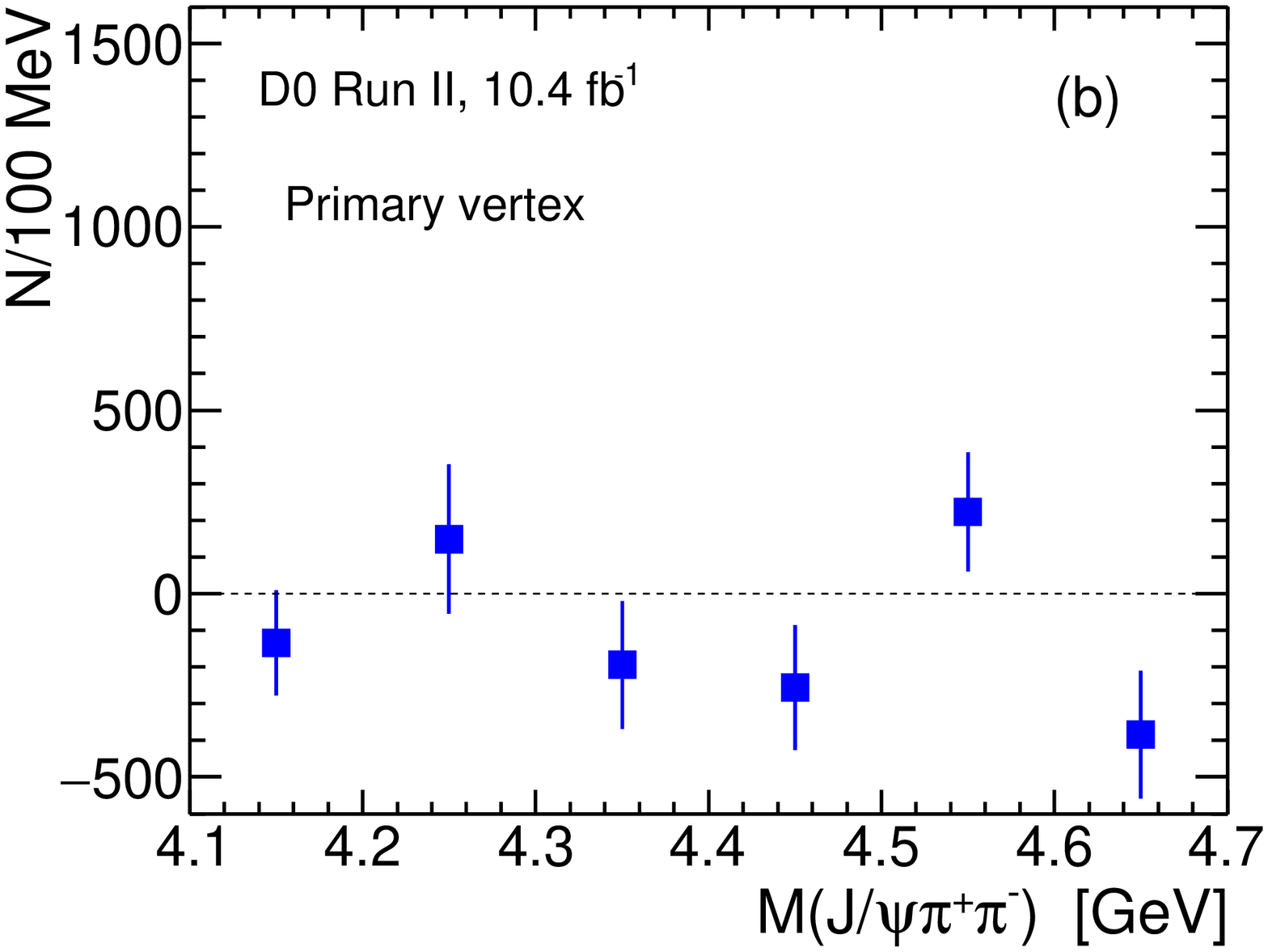}
\caption{\label{fig:zfromy} The $Z_c^+(3900)$ signal yield per 100 MeV
 for the six intervals of $m(J/\psi \pi^+ \pi^-)$:
4.1--4.2, 4.2--4.3, 4.3--4.4,
 4.4--4.5, 4.5--4.6 and 4.6--4.7~GeV
for (a) ``displaced vertex'' and (b) ``primary vertex'' selection. The points are placed at the bin centers.
}
\end{figure}

The  results of the fits are shown in Figs.~\ref{fig:npp1} and~\ref{fig:npp2}
and summarized in Table~\ref{tab:results} and in Fig.~\ref{fig:zfromy}.
The statistical significance of the signal 
is defined as $S=\sqrt{-2\, {\rm ln} ({\cal{L}}_0 /{\cal{L}}_{\rm max}) }$,
where ${\cal{L}}_{\rm max}$ and ${\cal{L}}_0$  are likelihood values at the
best-fit signal yield and the signal yield fixed to zero.
In the case of a negative signal yield, $S$ corresponds to the statisical
significance of the depletion.

For the ``displaced-vertex'' subsample we see a clear enhancement near
the $Z_c^{\pm}(3900)$  mass for events in the range $4.2<M(J/\psi \pi^+ \pi^-)<4.3$~GeV,
consistent with coming from the $\psi(4260)$ which has a mass of  $4230\pm8$~MeV~\cite{pdg},  
and a smaller excess in the ranges 4.5--4.6~GeV and  4.6--4.7~GeV.
In the mass interval  4.3--4.4~GeV (and to  smaller extent for  4.4--4.5~GeV)
our fits show a negative, but not significant, yield of $Z_c^{\pm}(3900)$ events.
There is no significant signal in the ``primary vertex'' subsamples in any  $M(J/\psi \pi^+ \pi^-)$
interval.

\begin{figure}[htb]
\includegraphics[scale=0.4]{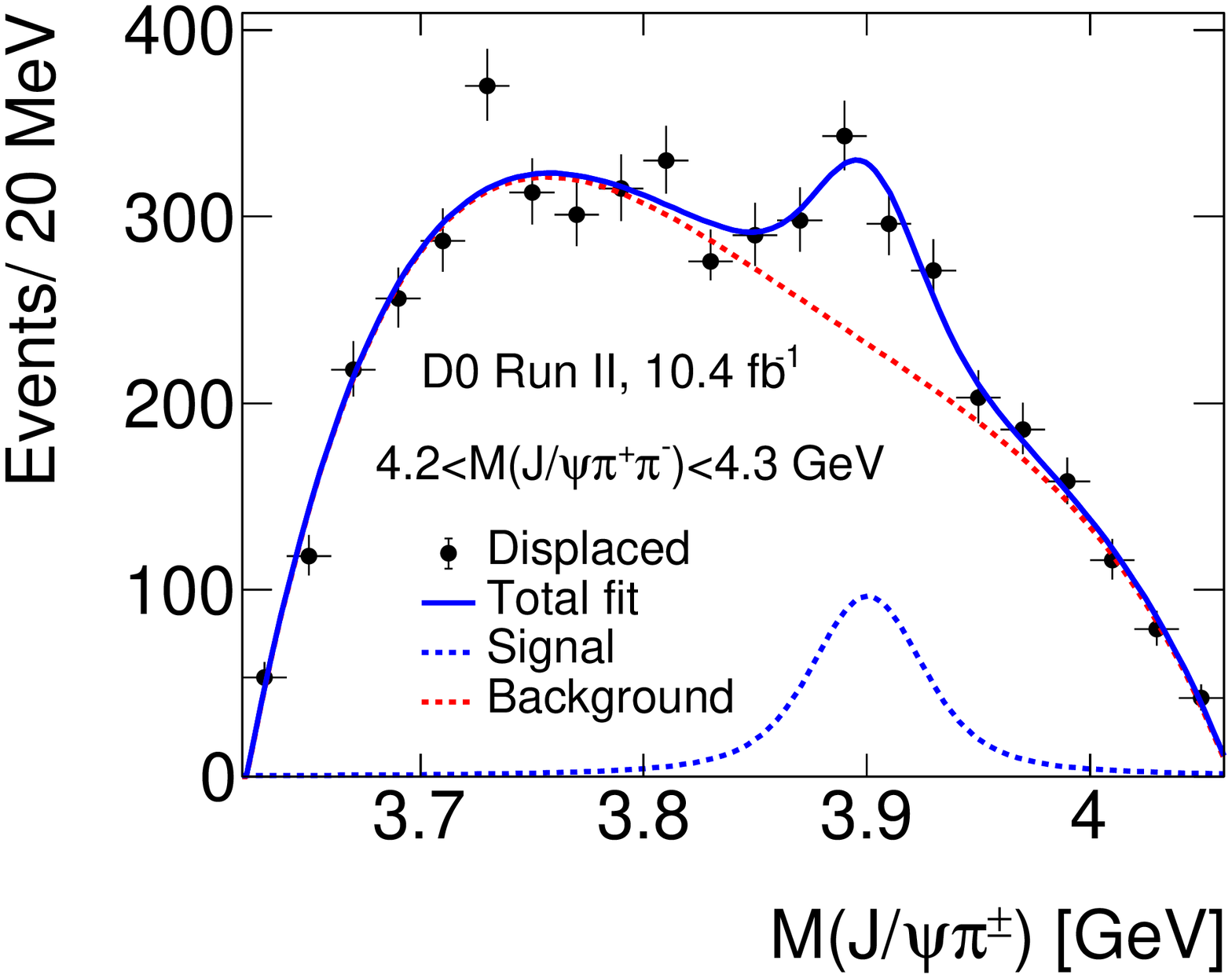}
\caption{\label{fig:freemg}
The  $J/\psi \pi^{\pm}$ invariant mass distribution  for the ``displaced-vertex''  candidates at
$4.2<M(J/\psi \pi^+ \pi^-)<4.3$~GeV.  The signal (solid blue line)  is modeled with
 a relativistic Breit-Wigner function
with free mass and width.  Background (dashed red line) is parametrized  as a 4th order Chebyshev polynomial.
}\end{figure}

For the ``displaced-vertex events'' in the mass range  $4.2<M(J/\psi \pi^+ \pi^-)<4.3$~GeV we also perform a fit
allowing  the signal mass and width to vary.
From this fit, shown in Fig.~\ref{fig:freemg},  we obtain our  best measurement of the $Z_c^{\pm}(3900)$ signal:
$M=3902.6 ^{+5.2}_{-5.0}$~MeV, $\Gamma=32 ^{+28}_{-21}$~MeV.
The signal yield is $N=364 \pm 156$  events, the fit quality is $\chi^2/ndf= 24.1/14$,
and the statistical significance is $S=5.4\sigma$.

%\begin{widetext}
\begin{table*}[htb]
\begin{center}
\caption{The $Z_c^{\pm}(3900)$ signal yields, fit quality, and statistical significance $S$ 
in intervals of $M(J/\psi \pi^+ \pi^-)$ for events with a displaced decay vertex and for the
complementary sample of ``primary vertex'' events, 
using the mass and width fixed at the PDG average values for
 the $J/\psi \pi^{\pm}$ channel: $M=3893.3$~MeV, $\Gamma=36.8$~MeV.}
\begin{ruledtabular}
\def\arraystretch{1.3}
\begin{tabular} {c|ccc|ccc} 
&& Displaced vertex &&& Primary vertex &\\
$M(J/\psi \pi^+ \pi^-)$ GeV & Event yield &  $\chi^2/ndf$ &  $S$ ($\sigma$)  & Event yield &  $\chi^2/ndf$ &  $S$ ($\sigma$) \\\hline 
4.1--4.2          & $86\pm 68$  &   18.7/14 & 1.3  & $-134\pm144$ & 52.7/15 & 0.9\\      
4.2--4.3          & $376\pm 76$ &   28.1/16 & 5.2  & $149\pm203$ & 21.9/14 & 0.5\\            
4.3--4.4          & $-148\pm 64$   &   17.4/15 & 2.3    & $194\pm174$ & 16.7/19 & 1.1   \\      
4.4--4.5          & $-33\pm60$   &   26.6/15 & 0.5    & $-256\pm170$ & 30.9/18 & 1.5  \\      
4.5--4.6          & $105\pm 64$ &   23.7/25 & 1.7  & $223\pm162$ & 42.3/23 & 1.4 \\ 
4.6--4.7          & $76\pm 55$ &   57.4/25 & 1.4  & $-384\pm174$ & 46.3/23 & 2.2 \\ 
%\hline \hline
\end{tabular}
\end{ruledtabular}
\label{tab:results}
\end{center}
\end{table*}
%\end{widetext}

%\clearpage

\section{\label{sec:eff}Acceptance of the displaced-vertex selection}

We obtain the acceptance of the ``displaced-vertex'' selection for $H_b$
decay events leading to $Z_c^{\pm}(3900)$
 using candidates for the
decay $B^0_d \rightarrow J/\psi K^{\pm}\pi^{\mp}$,
assuming that the distributions of the decay length and its uncertainty for  the $B^0_d$ decay
 are a good representation for the average $b$ hadron. 
Events are required to satisfy the same  kinematic and quality cuts as applied above.
We find  the fitted numbers of $B^0_d$ decays 
$N_{\rm displaced}=12951\pm167$ and  $N_{\rm primary}=6616\pm162$, respectively.
The ratios  of  $N_{\rm primary}$ to  $N_{\rm displaced}$ for
 $B^0_d$ and  $Z_c^{\pm}(3900)$ events with the same topology
should be the same,  to the extent
that the lifetimes of  $B^0_d$ and  $H_b$ are the same.  
With the  systematic uncertainty discussed in the next section taken into account,
the acceptance of the displaced vertex selection is $A =0.66\pm 0.02$.

\section{\label{sec:syst}Systematic uncertainties}
 
\subsection{Mass and width} 

We assign an asymmetric systematic uncertainty of $(0,+3)$~MeV to the mass measurement
due to a bias in mass measurements of $b$ hadrons at D0.
We assign the uncertainty on the mass and width due to uncertainty in the mass resolution
as half of the difference of the results obtained by changing the resolution by $\pm1\sigma$ 
to 15~MeV and 19~MeV.
 We assign uncertainties due to the background shape  based on the differences in the results
using the 3rd, 4th, and 5th-order polynomial.
The systematic uncertainties are summarized in Table~\ref{tab:systm}.

\begin{table}[htb]
\caption{\label{tab:systm} Systematic uncertainties in  the $Z_c^{\pm}(3900)$
mass and width measurements for Fig.~\ref{fig:freemg}. }
\begin{ruledtabular}
\def\arraystretch{1.3}
\begin{tabular}{lccc}
Source  & Mass, MeV   & Width, MeV &\\
\hline
Mass calibration & $^{+3}_{-0}$ & $0$ &\\
Mass resolution   & $\pm$0.1  &  $\pm 7$ &\\
Background shape & $\pm 1.4$ & $^{+25}_{-0}$ &  \\
\hline 
% & &   \\
Total (sum in quadrature) & $^{+3.3}_{-1.4}$ &$^{+26}_{-7}$ & \\
\end{tabular}
\end{ruledtabular}
\end{table}

\subsection{Signal  yields}

The uncertainty in the  relative yields of prompt and nonprompt production
of the $Z_c^{\pm}(3900)$ is dominated by statistical uncertainties.
The systematic uncertainties are evaluated as follows.
 
\begin{itemize}

\item Mass resolution

We assign the uncertainty in the signal yields  due to uncertainty in the mass resolution
as half of the difference of the results obtained by changing the resolution by $\pm1\sigma$ 
to 15~MeV and 19~MeV.

\item Trigger bias

Some of the single-muon triggers include a trigger term  requiring the presence of tracks
 with non-zero impact parameter. Events recorded solely by such triggers constitute 
approximately 5\% of all events.
% Assuming that such triggers are 100\% efficient for events in the displaced-vertex category and  reject ``primary vertex'' events, 
We assign a systematic uncertainty of $\pm5$\% to $N_{\rm displaced}$  due to this effect.

\item Acceptance of the displaced-vertex  selection

Our assumption of the equality of the displaced-vertex selection acceptance for
 the non-prompt $Z_c^{\pm}(3900)$
 and for $B^0_d$ is
based on expectation of the equality of the average  lifetime of  $b$-hadron parents of
 the $Z_c^{\pm}(3900)$ and that of the $B^0_d$. The world-average of the
$B^0_d$  lifetime is 3\% lower than the  lifetime averaged over all $b$ hadron species~\cite{pdg}.
This difference corresponds to a 1\% difference in the acceptance.  
In addition, there may be small differences between different channels in the transverse momentum distributions of the
parent $b$ hadrons and of the final-state particles. 
When the decay $B^0_s \rightarrow J/\psi \phi$ is used to estimate the ``displaced-vertex'' selection
acceptance,
the result is $A=0.675\pm0.010$.
We assign a 2\% uncertainty to the displaced-vertex acceptance  to account for
the differences between the $B^0_d$ decay and $H_b$ decays.

\item Signal model

We vary the fixed parameters~\cite{zcpdg1} of the signal mass and width by
$\pm$2.7~MeV and $\pm$6.5~MeV, respectively, corresponding to  $\pm$1$\sigma$.

\item Background shape

For the ``displaced vertex'' selection, we assign a symmetric uncertainty based on 
the differences between the results obtained using the 3rd, 4th, and 5th order polynomial.  
For the ``primary vertex'' selection, we assign an asymmetric uncertainty equal to the difference in the results
using the 5th-order and 4th-order polynomial.
The systematic uncertainties in the signal yield are summarized in Table~\ref{tab:systy}.

\end{itemize}

\begin{table}[htb]
\caption{\label{tab:systy} Systematic uncertainties in  the $Z_c^{\pm}(3900)$
signal yield for events in the  $4.2<M(J/\psi \pi^+ \pi^-)<4.3$~GeV interval (Fig.2c and 2d).}
\begin{ruledtabular}
\def\arraystretch{1.3}
\begin{tabular}{lccc}
Source & Displaced vertex  & Primary vertex &\\
\hline
Mass resolution   & $\pm$18  & $\pm$18  & \\
Trigger bias & $\pm$19  & -- &\\
Acceptance & $\pm$7   & -- & \\
Signal mass & $\pm$11 & $\pm 55$&\\
Signal width & $\pm40$  & $\pm 30$&\\
Background shape & $\pm$2 &   $^{+0}_{-149}$  &\\
\hline 
% & &   \\
Total (sum in quadrature) & $\pm49$   & $^{+65}_{-163}$ &\\
\end{tabular}
\end{ruledtabular}
\end{table}

\section{\label{sec:limit}Extracting limits on prompt  production rates}

Using results of the mass fits to the ``displaced-vertex'' and ``primary vertex''  subsamples
and the above value of  the acceptance of the displaced vertex selection, we can obtain
acceptance-corrected yields of prompt and nonprompt production 
and their ratio. We determine the yield  for the $J/\psi \pi^+ \pi^-$mass range 4.2--4.3~GeV  where
the nonprompt signal is statistically significant.

The mass spectrum  in the range 4.2--4.3~GeV 
in the ``primary vertex'' category shows no clear $Z_c^{\pm}(3900)$   signal and a large background of
about 5000$\pm$70 events in the  signal region. While there is no visible signal,
we cannot exclude a yield  comparable to the nonprompt signal.

In calculating the prompt-to-nonprompt ratio, we first obtain the total yield of the nonprompt
 production
 by dividing   $N_{\rm displaced}$ by the acceptance $A$.
That gives  $N_{\rm nonprompt}=570\pm 137$ (stat + syst).

Of the above number, a fraction equal to $1-A$ falls into the ``primary vertex'' category
 and must be subtracted
to obtain the net number of prompt events,  $N_{\rm prompt}=149-(1-0.66)\times 570=-45\pm 237$.
In calculating the uncertainty on the total prompt yield, we add  the statistical
and the systematic uncertainty components  in quadrature.
We obtain  the ratio 
 $r=N_{\rm prompt}/N_{\rm nonprompt}=-0.08^{+0.38}_{-0.46}$.
Assuming Gaussian uncertainties and setting the Bayesian prior for negative values of $r$ to zero,
 we obtain an upper limit of 0.70 at the  95\% credibility level.

\section{\label{sec:sum}Summary and conclusions}

Using the D0 Run II data reconstructed with a dedicated extended-tracking algorithm optimized
for low-$p_T$ tracks, we have studied production of the exotic state $Z^{\pm}_c(3900)$
in the  decays of $b$ hadrons to a $J/\psi \pi^+ \pi^-$ system with a subsequent
decay to $Z_c^{\pm}(3900) \pi^{\mp}$.
The observation is consistent with the sequential decay of a $b$-flavored hadron 
  $H_b \rightarrow \psi(4260) +{\rm anything}$,
 $\psi(4260) \rightarrow Z_c^{\pm}(3900)\pi^{\mp}$,  $Z_c^{\pm}(3900) \rightarrow J/\psi \pi^{\pm}$.  
We find a $Z_c^{\pm}(3900)$ signal at a  statistical significance of
5.4$\sigma$  for events
with   $4.2<M(J/\psi \pi^+ \pi^-)<4.3$~GeV, and find its mass and width
to be $M=3902.6 ^{+5.2}_{-5.0}{\rm \thinspace (stat)}^{+3.3}_{-1.4}{\rm \thinspace (syst)}$~MeV and
 $\Gamma=32 ^{+28}_{-21}{\rm \thinspace (stat)} ^{+26}_{-7}{\rm \thinspace (syst)} $~MeV
in agreement with world average values~\cite{pdg,zcpdg1}.

We searched for evidence of the prompt production of
 $\psi(4260)$ with subsequent rapid decays to  $Z_c^{\pm}(3900)\pi^{\mp}$.
In the absence
of a significant signal  we set an upper limit 
at the  95\% credibility level  on the ratio of prompt to nonprompt
production,  $N_{\rm prompt}/N_{\rm nonprompt}<0.70$.
This upper limit
 is significantly lower than that observed for $X(3872)$,
 for which $N_{\rm prompt}/N_{\rm nonprompt}$  is in the range two to three~\cite{7tev3872,8tev3872},
and $X(4140)$, for which  $N_{\rm prompt}/N_{\rm nonprompt}$$\approx1.5$~\cite{d04140}.

% acknowledgement_APS_full_names.tex             18 June 2018
%
% Acknowledgement paragraph in English of Oct. 15, 2014 for APS journals

This document was prepared by the D0 collaboration using the resources of the Fermi National Accelerator Laboratory (Fermilab),
a U.S. Department of Energy, Office of Science, HEP User Facility. Fermilab is managed by Fermi Research Alliance, LLC (FRA),
acting under Contract No. DE-AC02-07CH11359.

We thank the staffs at Fermilab and collaborating institutions,
and acknowledge support from the
Department of Energy and National Science Foundation (United States of America);
Alternative Energies and Atomic Energy Commission and
National Center for Scientific Research/National Institute of Nuclear and Particle Physics  (France);
Ministry of Education and Science of the Russian Federation, 
National Research Center ``Kurchatov Institute" of the Russian Federation, and 
Russian Foundation for Basic Research  (Russia);
National Council for the Development of Science and Technology and
Carlos Chagas Filho Foundation for the Support of Research in the State of Rio de Janeiro (Brazil);
Department of Atomic Energy and Department of Science and Technology (India);
Administrative Department of Science, Technology and Innovation (Colombia);
National Council of Science and Technology (Mexico);
National Research Foundation of Korea (Korea);
Foundation for Fundamental Research on Matter (The Netherlands);
Science and Technology Facilities Council and The Royal Society (United Kingdom);
Ministry of Education, Youth and Sports (Czech Republic);
Bundesministerium f\"{u}r Bildung und Forschung (Federal Ministry of Education and Research) and 
Deutsche Forschungsgemeinschaft (German Research Foundation) (Germany);
Science Foundation Ireland (Ireland);
Swedish Research Council (Sweden);
China Academy of Sciences and National Natural Science Foundation of China (China);
and
Ministry of Education and Science of Ukraine (Ukraine).


\begin{thebibliography}{99}
\bibitem{ip}
 The impact parameter  {\sl IP}  is defined as the distance of closest approach of the track to the $p\bar{p}$
 collision point projected onto the plane transverse to the $p\bar{p}$ beams.
\bibitem{eta}
$\eta=-\ln[\tan(\theta/2)]\ $ is the pseudorapidity and $\theta$ is
the polar angle between the track momentum and the proton beam direction. $\phi$ is the azimuthal angle 
of the track.
\bibitem{zcpdg1} Ref.~\cite{pdg} lists the $Z_c(3900)$ as a two-channel resonance
and quotes average mass for the decays $Z_c(3900) \rightarrow D \bar D^*$ and
$Z_c(3900)^{\pm,0} \rightarrow J/\psi \pi^{\pm,0}$. The mass values measured in
the two channels differ by 11~MeV and  the compatibility
is $\approx$4$\times 10^{-6}$. We choose to calculate the average mass and width
for the $J/\psi \pi^{\pm}$ channel only, using the PDG prescription for measurement
averaging.
\end{thebibliography}
\end{document}